# Shear thickening in electrically stabilized non-aqueous colloidal suspensions

Joachim Kaldasch* and Bernhard Senge

Technische Universität Berlin
Fakultät III: Lebensmittelrheologie
Königin-Luise-Strasse 22
14195 Berlin
Germany

Jozua Laven

Eindhoven University of Technology
Laboratory of Materials and Interface Chemistry
PO Box 513
5600 MB Eindhoven
The Netherlands

* author to who correspondence should be sent

**Abstract**

The authors previously introduced an activation model for the onset of shear thickening in electrically stabilized colloidal suspensions. It predicts that shear thickening occurs, when particles arranged along the compression axis in a sheared suspension do overcome the electrostatic repulsion at a critical shear stress, and are captured in the primary minimum of the DLVO interaction potential. A comparison with an experimental investigation on non-aqueous silica suspensions, carried out by Maranzano and Wagner, is performed. For particle systems that fall into the applicability range of the theory, a good coincidence between the experimental data and the model predictions can be found.

# 1. Introduction

A number of theoretical attempts were made in the past to explain shear thickening in electrically stabilized colloidal suspensions. There are four main research directions. One is to neglect Van der Waals attraction between colloidal particles and consider them as effective hard spheres. For this type of interaction substantial progress in the understanding of shear thickening has been achieved by applying computer simulations. The idea is that short-range lubrication forces are responsible for the formation of large shear induced density fluctuations, known as hydroclusters [1,2,3]. A second approach is to relate shear thickening to an order-disorder transition, where an ordered, layered structure becomes unstable above a critical shear rate [4,5,6]. A third direction is based on a mode-coupling model, where the memory term takes into account the density, the shear stress and the shear rate. It was suggested that shear thickening can be understood as a stress-induced transition into a jammed state [7,8,9]. Recently an activation model taking into account the total interaction potential was presented [10].

The key idea of this paper is to apply this activation model to electrically stabilized suspensions in non-aqueous media. We focus on such media because in such a case electrical interactions can be expected to be well described by the Derjaguin-Landau-Verwey-Overbeek (DLVO) theory, while any complications that might arise from aqueous non-DLVO effects (e.g., hydration and hydrophobic interactions) are avoided. As schematically displayed in Fig.1, the total DLVO interaction potential, $U(h)$, as a function of the surface-to-surface distance $h$, consists of a primary and a secondary minimum due to a combined action of Van der Waals attraction and electrostatic repulsion. The model suggests that reversible shear thickening occurs, when particles along the compression axis are forced by an applied shear stress to overcome their mutual repulsion. In the case of a deep primary minimum, however, irreversible coagulation prohibits a disruption of the particles by shear flow, and the model describes the initial states of an orthokinetic coagulation [11,12,13].

In the next section the activation model is summarized, followed by a comparison of the model predictions with an experimental investigation on non-aqueous suspensions by Maranzano and Wagner [14], where non-DLVO forces (hydration and hydrophobic forces) can be neglected.

## 2. The Model

We want to consider a concentrated, electrically stabilized, colloidal suspension of monodisperse spherical particles, and suppose, that the particle interaction can be described by a superposition of an electrostatic repulsion $U_{el}(h)$, and Van der Waals attraction $U_{vdW}(h)$:

$$U(h) = U_{el}(h) + U_{vdW}(h) \tag{1}$$

The electrostatic repulsion of the particles is due to the overlap of their electric double layers. We confine our discussion to particles that have a constant, i.e. distance-independent, surface potential $\Psi_0$ or surface charge density. In numerous cases a suitable approximation of the electrostatic repulsion for a constant surface potential approach of two particles can be written as [15]:

$$U_{CP}(h) = 2\pi \varepsilon_0 \varepsilon_r a \varsigma^2 \ln(1 + e^{-\kappa h}) \tag{2}$$

and for a constant surface charge approach:

$$U_{CC}(h) = -2\pi \varepsilon_0 \varepsilon_r a \varsigma^2 \ln(1 - e^{-\kappa h}) \tag{3}$$

where we approximated the surface potential $\Psi_0$ by the zeta potential $\zeta$. The parameter $\varepsilon_0$ is the absolute and $\varepsilon_r$ the relative dielectric constant, while $a$ is the particle radius. The Debye reciprocal length $\kappa$ is defined by:

$$\kappa = \sqrt{\frac{2 C_S N_A z^2 e_0^2}{\varepsilon_0 \varepsilon_r k_B T}} \tag{4}$$

where $e_0$ is the elementary electric charge, $N_A$ the Avogadro- number, $z$ the ionic charge number and $C_S$ is the salt concentration.

The non-retarded Van der Waals attraction between two spheres can be taken into account by

$$U_{vdW}(h) = -\frac{Aa}{12h} \tag{5}$$

where $A$ is the effective Hamaker constant determined by the dielectric constants of the solvent-particle combination.

Applying a simple shear, along the compression axis of the sheared suspension, colloidal particles may overcome their repulsive potential barrier. The energy barrier formed by the interaction potential is given by:

$$U_B = U(h_{max}) - U(h_0)$$

(6)

where $h_{max}$ (see Fig.1) is the gap at the maximum of the interaction potential determined by

$$\frac{\partial U(h)}{\partial h} = 0; \frac{\partial^2 U(h)}{\partial h^2} < 0$$

(7)

The parameter $h_0$ is related to the surface-to-surface distance at volume fraction $\Phi$ according to:

$$h_0(\Phi) \cong 2a\left(\left[\frac{\Phi_m}{\Phi}\right]^{1/3} - 1\right)$$

(8)

where $\Phi_m=0.64$ is the maximum random packing density of a hard sphere suspension.
 Near the isoelectric point (IEP), which corresponds to zero $\zeta$-potential, the electrostatic repulsion between the colloidal particles can be neglected. In this case unstable particles grow to clusters by thermal collisions [11]. Increasing the repulsion, the diffusion controlled coagulation, known as perikinetic coagulation, is slowed down. The perikinetic coagulation becomes an activated process characterized by the stability ratio:

$$W = \frac{N}{N_C}$$

(9)

where $N$ is the number of thermal collisions and $N_C$ is the number of collisions leading to coagulation. The stability ratio can be approximated by [16]:

$$W \cong \frac{1}{2\kappa a}\exp\left(\frac{U(h_{max})}{k_B T}\right)$$

(10)

where $k_B$ is the Boltzmann constant and $T$ the temperature. A suspension with a stability ratio of $W \approx 10^{10}$ usually takes months to coagulate. Since the present model is applicable only for stable colloidal particles, we want to demand that $W \geq 10^{10}$.
 An applied shear stress creates a bias potential $U_S$ along the compression axis:

$$U_S = \sigma V^*$$

(11)

Pushing two particles together the activation volume $V^*$ is of the order of the free volume per particle:

$$V^* = V_0 \frac{\Phi_m}{\Phi}$$
(12)

The frequency of a shear induced (activated) coagulation is of the order $f \sim exp(-(U_B-U_S)/k_BT)$. Shear thickening can be expected to occur, when $U_B=U_S$. The critical stress becomes [10]:

$$\sigma_C = \frac{U_B}{V^*}$$
(13)

For small shear rates the electrostatic repulsion between two approaching particles can be described by equilibrium formulas as given in Eqs. 2 and 3 for the constant potential and constant charge approximation. However with increasing shear rates the approaching time of the particles may be comparable with the relaxation time of the electric double layer. Between two approaching spheres the characteristic convective time of the double layer at the transition is given by $\tau_C = (v\kappa)^{-1}$. The approaching velocity $v$ of the particles along the compression axis can be estimated by equating the external mean field force on a particle with the viscous resistance of two approaching particles:

$$\sigma_C \pi a^2 = 6\pi\eta_S a \frac{a}{h} v$$
(14)

where $\eta_S$ is the solvent viscosity. Approximating $h \sim \kappa^{-1}$, we find for the convective time scale at the critical stress:

$$\tau_C \sim \frac{6\eta_S}{\sigma_C}$$
(15)

The diffusive time scale for the lateral relaxation of the double layer due to the squeezing of the gap between two approaching spheres can be estimated by:

$$\tau_D \sim \frac{(2a)^2}{D}$$
(16)

Here D is the diffusion constant given by

$$D = \frac{k_B T}{6\pi\eta_S a_i}$$
(17)

where $a_i$ is the radius of the ions. The corresponding Péclet number can be written as the ratio between the convective and the diffusive time scales:

$$Pe = \frac{\tau_C}{\tau_D} = \frac{6\eta_S D}{4\sigma_C a^2}$$

(18)

It is known that particles that have a constant potential (CP) behavior at equilibrium adjust their surface charge when they approach each other, in order to keep the surface potential constant. This requires ions to diffuse from/to the surface. When *Pe* goes down to *Pe=O(1)* there is not enough time for the surface to fully adapt the number of charges at the surfaces bordering the gap between the particles in such a way that a constant surface potential is preserved. In fact the surfaces rather tend to keep their charge density constant. Although the corresponding diffuse double layer in such dynamical situation is not exactly identical to that of the equilibrium constant charge (CC) case, we suppose that Eq. 3 will not be far off the reality.

The critical stress that corresponds to such transition is given by

$$\sigma_C^* = \frac{6\eta_S D}{4a^2}$$

(19)

Since an electric double layer may respond according to the CP or the CC case, the activation barrier can be evaluated by setting $U_{el} = U_{CP}$ for a CP and $U_{el} = U_{CC}$ for a CC double layer response. The corresponding critical stresses are denoted as $\sigma_C^{CP}$ and $\sigma_C^{CC}$.

## 3. Comparison with Experimental Investigations

We want to compare the present model with experimental results on non-aqueous suspensions, and focus here on an extensive experimental investigation carried out by Maranzano and Wagner [14]. They studied electrically stabilized silica particles in THFFA, in order to minimize the v. d. Waals attraction. Since the refractive index of the silica particles matches the solvent fluid, the Hamaker constant can be estimated from [17]:

$$A \cong \frac{3k_B T}{4}\left(\frac{\varepsilon_{THFFA} - \varepsilon_S}{\varepsilon_{THFFA} + \varepsilon_S}\right)^2$$

(20)

Using $\varepsilon_S = 4.5$ (silica) and $\varepsilon_{THFFA} = 8.2$ for the dielectric constants of this particle-solvent system the Hamaker constant becomes $A \cong 0.41 \ast 10^{-21}$ J, while the radius of the silanol groups are estimated to be of the order $a_i = 0.3$ nm.

Maranzano and Wagner investigated suspensions of five different particle diameters at a number of volume fractions. The characteristic data are summarized in Table 1. In order to learn in what cases the present model might be applicable to Maranzano's systems, the stability of these systems as judged from the stability ratio $W$ in Eq.(10) can be checked in the last column of Table 1. Based on the remarks made in Section 2, it can be concluded that the activation model can be applied with high confidence only to the samples HS150, HS600 and HS1000. Thus, the systems HS300 and HS75 will not be analyzed here.

Figures 2-4 display the experimental results of the samples HS1000, HS600 and, HS150 together with the theoretically expected critical shear stresses ($\sigma_C^{CC}$, $\sigma_C^{CP}$) for the onset of shear thickening (solid lines). The error bars indicate the uncertainty of the experimental data due to the finite resolution of the measurement. Also displayed is the critical stress $\sigma_C^*$ which represents the border between the regimes of constant potential and constant charge. While for $\sigma_C < \sigma_C^*$, the onset of shear thickening can be expected to take place at CP particle interaction, for $\sigma_C > \sigma_C^*$ the model suggests that onset to occur at CC conditions. Note that it cannot be determined a priori whether $\sigma_C <> \sigma_C^*$.

For the HS1000 samples the experimental critical stresses fulfill the condition $\sigma_C < \sigma_C^*$, and are therefore expected to be given by $\sigma_C^{CP}$. Although not all data points match the predicted critical stress they are arranged close to $\sigma_C^{CP}$. On the other hand, the experimental data of the HS600 and HS150 system match the critical shear stress predictions $\sigma_C^{CC}$, in accordance with the fact that $\sigma_C > \sigma_C^*$. The data point of sample HS150 at $\Phi = 0.44$ cannot be explained by the present model. We interpret it as an outlier.

For comparison the predictions of the critical stress $\sigma_C^{MW}$ for the onset of shear thickening as based on an effective hard sphere model by Maranzano and Wagner, which neglects Van der Waals attraction, are evaluated according to [14,18]:

$$\sigma_C^{MW} = 0.096 \frac{\pi \kappa \varepsilon_0 \varepsilon_r \Psi_0^2}{a}$$

(21)

In Figures 2-4 the critical stresses $\sigma_C^{MW}$ are displayed, indicated by a dotted line. The effective hard sphere model appears usually to overestimate the critical stresses, by up to a factor 20 (with system HS1000), and demonstrates that ignoring the Van der Waals attraction is not justified. We want to emphasize that the experimental determination of the critical stresses is difficult, in particular when the transition into shear thickening occurs smoothly.

Although not all experimental data points could be predicted by the present version of the theory, it gives a qualitative better picture of the shear thickening instability than the effective hard sphere model by Maranzano and Wagner.

## 4. Conclusion

The present investigation reveals that, within the applicability range, the activation model of shear thickening is in good agreement with experimental data of the critical stress obtained from electrically stabilized non-aqueous suspensions. This result is more satisfactory than what was previously found for aqueous suspensions where only the proper order of magnitude of the critical stress could be given. For practical applications, the present model delivers a good indication of the critical shear stresses for the onset of shear thickening in electrically stabilized non-aqueous suspensions.

As shown in Table1 the maximum of the interaction potential, $h_{max}$, is at distances of a few hundred *pm* from the surface. Since a silicon atomic diameter is ~200 pm, the roughness of the particle surface prohibits a direct contact between the particles beyond the maximum of the interaction potential. We therefore suggest that the transition between reversible shear thickening and irreversible orthokinetic coagulation is smooth, and essentially determined by the distance of the maximum of the interaction potential to the surface and by the surface roughness of the particles. It is quite surprising that the presented continuum model gives that good predictions of the critical stress.

A shortcoming of the model is that it is unable to predict shear thickening in sterically stabilized suspensions. Unlike the interaction potential predicted by the DLVO-theory, which shows two minima and one maximum (Fig.1), sterically stabilized systems exhibit a potential, which consists only of a shallow attractive minimum followed by a rapid increase as the surfaces of the colloidal particles approach each other closely. The absence of a potential maximum implies that these particle systems cannot be treated with the activation model, while the effective hard sphere model should rather be applicable.

We have to emphasize that present computer simulations on electrically stabilized suspensions usually ignore the double layer dynamics and the surface roughness of the particles. Therefore these simulations are not accurate enough to be comparable with real systems. Simulations taking into account both effects are necessary in order to improve our understanding of the shear thickening phenomenon.

- 9 -**Tables**

| Suspension | a [nm] | ζ [mV] | κa | $h_{max}$ [pm] | W |
|---|---|---|---|---|---|
| HS 75 | 37.5 | 42.6 | 13.2 | 680 | 0.1 |
| HS 150 | 83.5 | 72.1 | 28.5 | 237 | $1*10^{11}$ |
| HS 300 | 151 | 42.3 | 41.7 | 450 | $4*10^5$ |
| HS 600 | 304 | 92.7 | 87.1 | 200 | $9*10^{77}$ |
| HS1000 | 328 | 68.2 | 89.9 | 280 | $8*10^{42}$ |

**Table 1.** Characteristic parameters of the silica particles in THFFA suspensions investigated by Maranzano and Wagner (2001a), where $h_{max}$ and $W$ are determined at constant potential.. We used $k_B=1.38\ 10^{-23}$ J, $e=1.6\ 10^{-19}$ As, $\varepsilon_r=8.2$, $\varepsilon_0=8.854\ 10^{-12}$ F/m, $z=1$, $T=298$ K, $\eta_{THFFA}=0.005$ Pas.

**Figures**

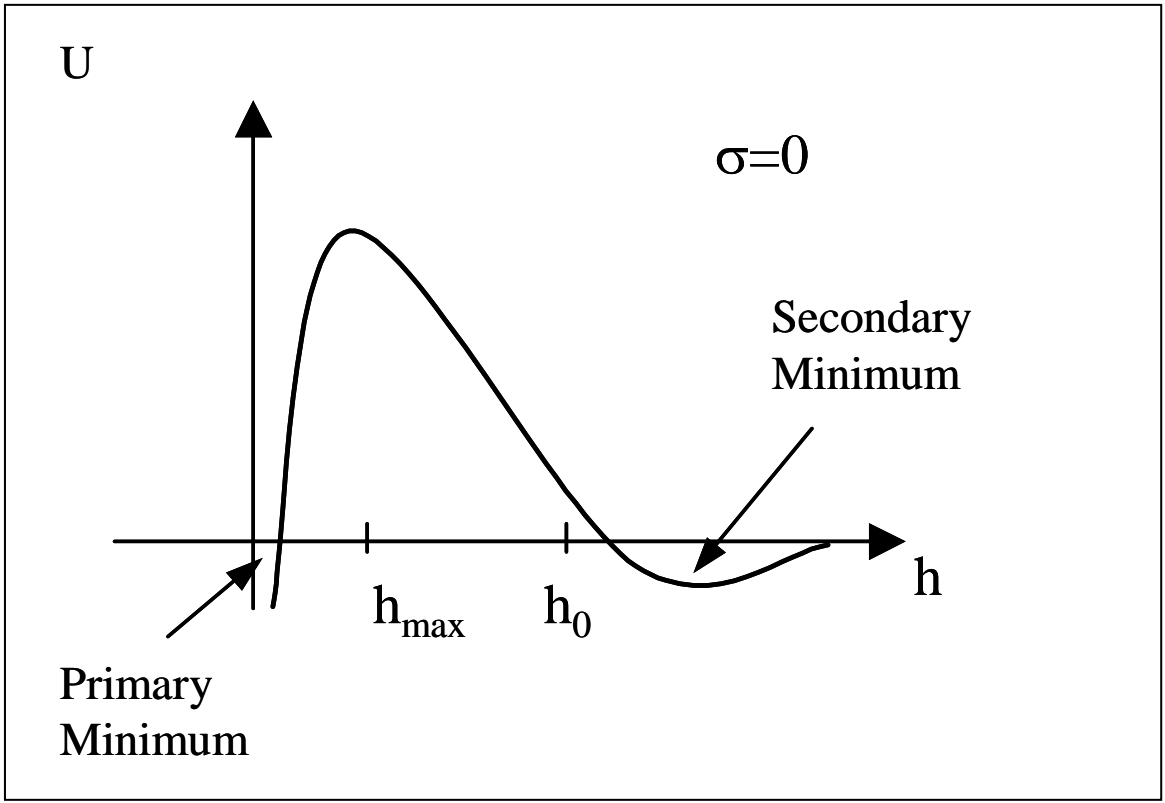

**Figure 1.** Schematic representation of the DLVO interaction potential *U* as a function of the two-particle surface-to-surface distance *h*.

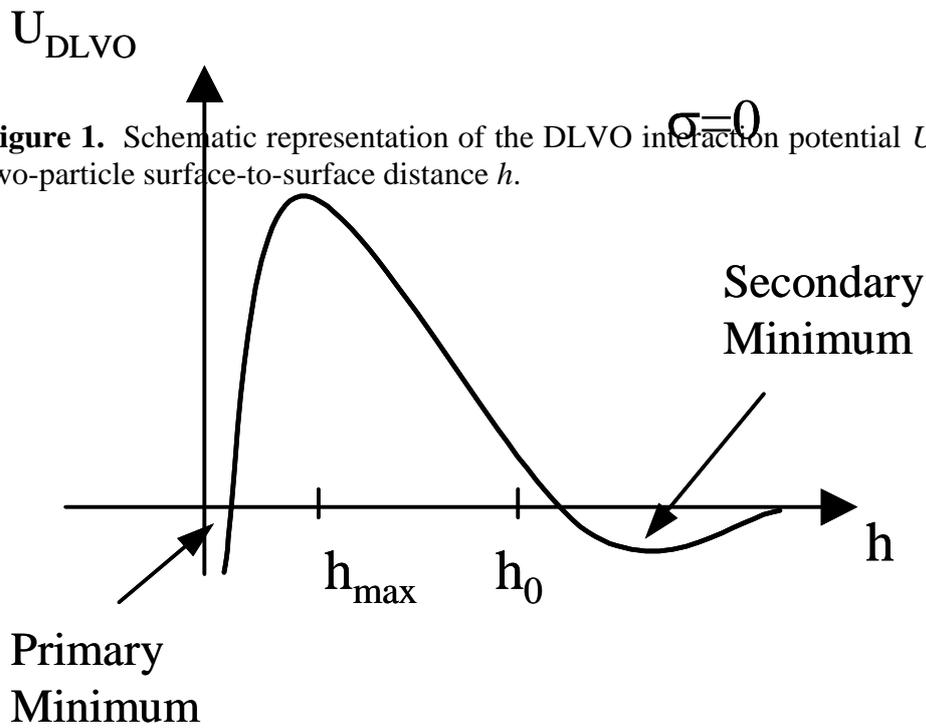

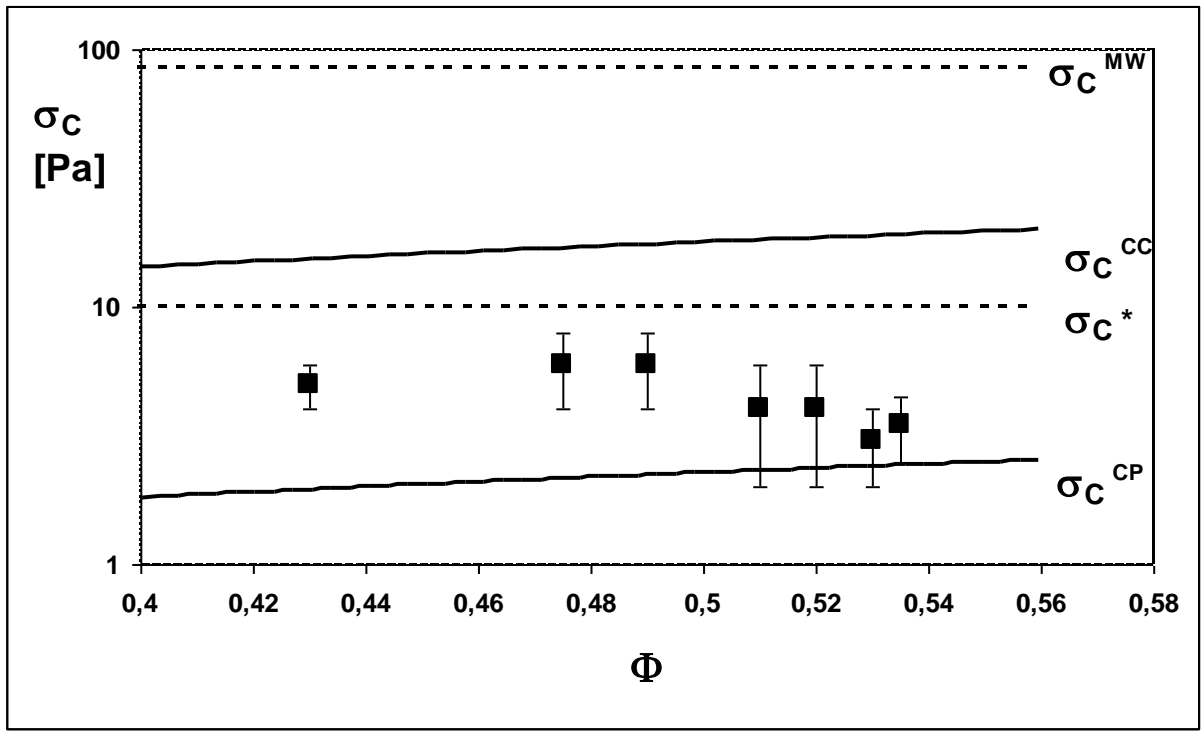

**Figure 2.** The critical shear stress at the onset of shear thickening of the samples HS 1000 at different volume fractions [14]. Because $\sigma_C < \sigma_C^*$, the critical stress can be expected to be given by $\sigma_C^{CP}$.

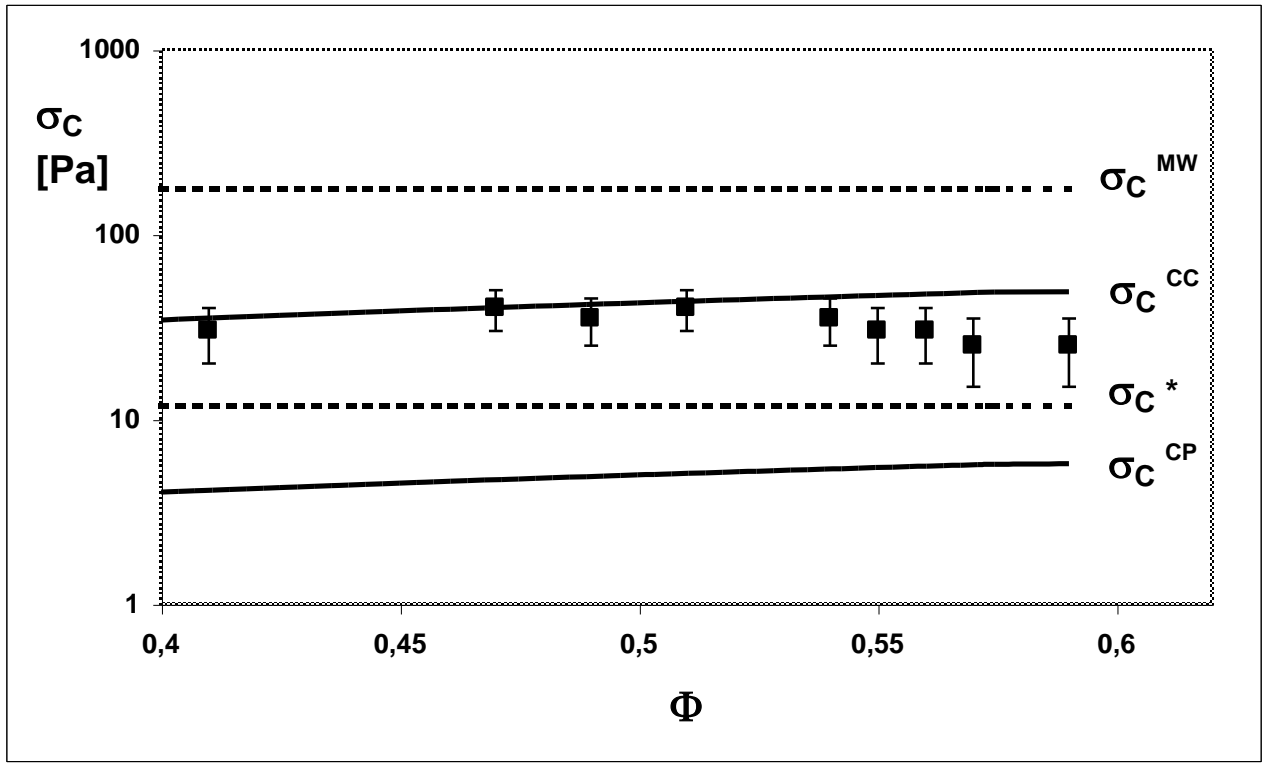

**Figure 3.** The critical shear stress at the onset of shear thickening of the sample HS 600 at different volume fractions [14]. Because $\sigma_C > \sigma_C^*$, the critical stress can be expected to be given by $\sigma_C^{CC}$.

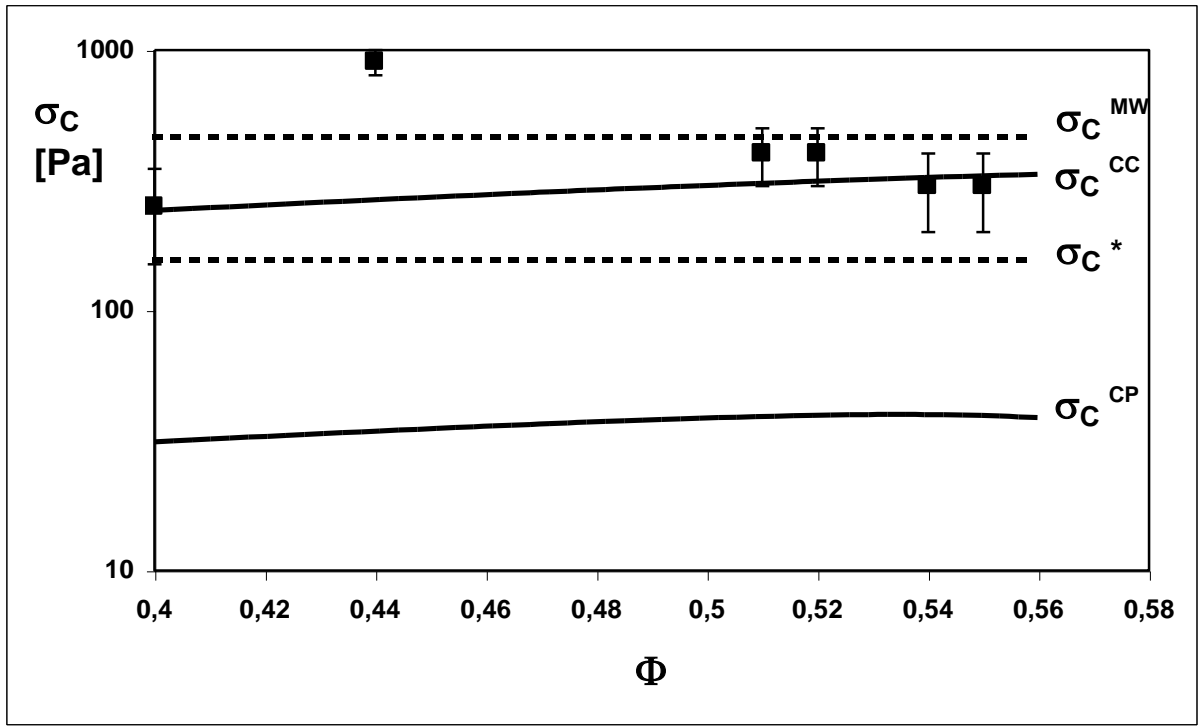

**Figure 4.** The critical shear stress at the onset of shear thickening of the samples HS 150 at different volume fractions [14]. Because $\sigma_C > \sigma_C^*$, the critical stress can be expected to be given by $\sigma_C^{CC}$.